\documentclass{aa}    

\usepackage{txfonts}
\usepackage{longtable}
\usepackage{natbib}
\usepackage{graphicx}

\bibpunct[, ]{(}{)}{;}{a}{}{,}
\usepackage{supertabular}

\usepackage{color}

\newcommand{\SampleStar}{HE~1405$-$0822}        
\newcommand{\eps}[1]{\log\varepsilon_{\rm #1}}

\begin{document}

\title{The Hamburg/ESO R-process Enhanced Star survey (HERES) \thanks{Based on
    observations collected at the European Southern Observatory, Paranal,
    Chile (Proposal numbers 170.D-0010G, and 170.D-0010J).}}
\subtitle{VIII. The r+s star HE~1405$-$0822}

\author{
  W.Y. Cui\inst{1,2},
  T. Sivarani\inst{3}, \and
  N. Christlieb\inst{1}
}

\offprints{W.Y. Cui; \email{cui@lsw.uni-heidelberg.de}}
\institute{Zentrum f\"ur Astronomie der Universit\"at Heidelberg, Landessternwarte,
     K\"onigstuhl 12, 69117 Heidelberg, Germany
     \\ \email{cui@lsw.uni-heidelberg.de}
\and Department of Physics, Hebei Normal University, Nanerhuan dong
Road 20, D-050024 Shijiazhuang, China
     \\ \email{cuiwenyuan@hebtu.edu.cn}
\and Indian Institute of Astrophysics, Bangalore, India\\
     \email{sivarani@gmail.com}
\and Zentrum f\"ur Astronomie der Universit\"at Heidelberg, Landessternwarte,
     K\"onigstuhl 12, 69117 Heidelberg, Germany \\
     \email{N.Christlieb@lsw.uni-heidelberg.de}
}

\date{Received  / Accepted }

\abstract{}{
  The aim of this study is a detailed abundance analysis of the newly
  discovered r-rich star HE~1405$-$0822, which has
  $\mathrm{[Fe/H]}=-2.40$. This star shows enhancements of both r- and s-elements ,
  $\mathrm{[Ba/Fe]}=+1.95$ and
  $\mathrm{[Eu/Fe]}=1.54$, for which reason it is called r+s star.
}{
  Stellar parameters and element abundances were determined by
  analying high-quality VLT/UVES spectra. We used Fe I line
  excitation equilibria to derive the effective temperature.
  The surface gravity was
  calculated from the \ion{Fe}{i}/\ion{Fe}{ii} and
  \ion{Ti}{i}/\ion{Ti}{ii} equilibria.
}{
  We determined accurate abundances for 39 elements, including 19
  neutron-capture elements. HE~1405$-$0822 is a red giant. Its
  strong enhancements of C, N, and s-elements are the consequence of enrichment by a former
  AGB companion with an initial mass of less than
  $3\,\mathrm{M}_\odot$. The heavy n-capture element abundances
  (including Eu, Yb, and Hf) seen in HE~1405$-$0822 do not agree
  with the r-process pattern seen in strongly r-process-enhanced
  stars. We discuss possible enrichment scenarios for this star. The
  enhanced $\alpha$ elements can be explained as the result of enrichment by
  supernovae of type II. Na and Mg may have partly been synthesized in a
  former AGB companion, when the primary $^{22}$Ne acted as a
  neutron poison in the $^{13}$C-pocket.
}
{}


\titlerunning{Abundance analysis of HE~1405$-$0822}
\authorrunning{Cui et al.}

\maketitle
%
\section{Introduction}\label{Sect:intro}

Elements beyond the iron group are believed to be mostly synthesized
through neutron-capture (n-capture hereafter) processes, which
consist of the rapid (r-) and the slow (s-) process. These are
distinguished by the timescales for neutron captures relative to the
$\beta$-decay timescales of the resulting nuclei
\citep{Burbidge1957RvMP}.  The s-process is generally assumed to
take place in the asymptotic giant branch (AGB) phase of stars of
low or intermediate mass.  However, \citet{Pignatari2008nuco.conf}
suggested that in metal-poor, fast-rotating stars (hereafter spin
stars), the efficiency of the s-process is high enough to produce
the strong overabundances of Sr, Y, and Zr observed in extremely
metal-poor halo stars ($\mathrm{[Fe/H]} <-3.0$)\footnote{The
standard spectroscopic notation
  is used, i.e., $\mathrm{[X/H]} =\log_{10}(N_{\rm X}/N_{\rm H})_{\star}
  - \log_{10}(N_{\rm X}/N_{\rm H})_{\odot}$, where $N_{\rm X}$ is the
  number density of atoms of the element X.}, where the AGB stars do not
have time to contribute. Using models of extremely metal-poor spin
stars, \citet{Chiappini2011Natur} successfully explained the large
scatter in [Y/Ba] ratios in NGC~6522, the oldest globular cluster of
the Milky Way. This large scatter is also seen in the most
metal-poor halo stars.

Some explosive astrophysical events are usually accompanied by
synthesis of the r-process elements, but the exact site(s) of this
process is still unclear \citep{Sneden2008ARA&A}. Several
possibilities have been suggested, including prompt explosions of
core-collapse (Type II/Ibc) supernovae \citep{Wanajo2003ApJ},
neutron star mergers
\citep{Lattimer1977ApJ,Rosswog1999A&A,Freiburghaus1999ApJ},
neutrino-driven winds \citep{Woosley1994ApJ,Wanajo2001ApJ}, the
accretion-induced collapse mechanism
\citep[AIC,][]{Qian2003ApJ,Cohen2003ApJ}, and Type 1.5 supernovae
\citep{Iben1983ARA&A,Zijlstra2004MNRAS}. More observational studies
of r-process-enriched stars may shed light on this research field.

Recent studies indicate that there may be two separate r-processes,
which are referred to as the main r-process, which is responsible
for the creation of heavy n-capture elements with $Z\ge56$
\citep{Truran2002PASP,Sneden2003ApJ}, and a weak r-process for the
light n-capture elements with $Z<56$
\citep{Kratz2007ApJ,Wanajo2006NuPhA}.  In the strongly r-process
enhanced stars, i.e. stars with $\mathrm{[Eu/Fe]}>+1.0$ and
$\mathrm{[Ba/Eu]}<0$, \citep[hereafter r-II
stars,][]{Beers2005ARA&A}, the abundance distribution of heavy
n-capture elements does not vary significantly from star to star,
and it agrees very well with the scaled solar system r-process
distribution.  Up to now, about ten r-II stars have been discovered,
including CS 22892$-$052 \citep{Sneden2003ApJ,Sneden2009ApJS},
CS\,31082$-$001 \citep{Hill2002A&A}, CS\,31078-018
\cite{Lai2008ApJ}, HD~221170 \citep{Ivans2006ApJ,Sneden2009ApJS},
BD$+17^\circ\,3248$
\citep{Cowan2002ApJ,Cowan2011rrls.conf,Roederer2010aApJ}, CS
22953$-$003 \citep{Francois2007A&A}, HE~1523$-$0901
\citep{Frebel2007ApJ}, HD~115444
\citep{Westin2000ApJ,Sneden2009ApJS,Hansen2011A&A}, CS\,29491$-$069
\citep{Hayek2009A&A}, HE~1219$-$0312 \citep{Hayek2009A&A}, and
HE~2327$-$5642 \citep{Mashonkina2010A&A}. Studies on these r-II
stars confirmed the universal pattern of the main r-process
\citep{Cowan2011rrls.conf}.

Unlike the main r-process pattern seen in r-II stars, there are
significant deviations between the light n-capture elements ($37\le
Z\le47$, i.e., from Rb to Ag) in r-II stars and the scaled solar
system r-process pattern.  This implies that multiple r-process
sites \citep{Wasserburg1996ApJ, Wasserburg2000ApJ,
Qian2000PhR,Qian2001ApJ,Qian2002ApJ} or the same core-collapse
supernovae but different epochs or regions
\citep{Cameron2001ApJ,Cameron2003ApJ} are probably responsible for
the solar r-process distribution. Some r-process-poor stars, such as
HD~122563 and HD~88609, show a high excess of light n-capture
elements (e.g., Sr, Y and Zr), but no enrichment of the heavy ones
(e.g., Ba, Eu), which indicates that the weak r-process plays a
dominant role in producing their abundance patterns
\citep{Honda2006ApJ,Honda2007ApJ,Izutani2009ApJ}. In fact, a
combination of processes, such as the weak r-process and the main
r-process, is more efficient in reproducing the observed abundances
of light n-capture elements for many such types of stars
\citep{Zhang2010MNRAS,Roederer2010aApJ,Roederer2010bApJ,Arcones2011ApJ,Cowan2011rrls.conf}.
Though the n-capture element distribution in CS 22892$-$052 is not
similar to that of most metal-poor stars, \citet{Cowan2011rrls.conf}
points out that the r-process enrichment in the early Galaxy is
common because of the presence of Sr, Ba, etc. in nearly all
metal-poor stars that do not show s-process enhancements.

Some metal-poor s-rich stars at the same time show a strong
enrichment of Eu and other heavy neutron-capture elements, which in
the solar system are predominantly produced by the r-process. These
stars are commonly referred to as r$+$s stars
\citep{Beers2005ARA&A}. Their s-process enrichment is usually
attributed to mass transfer (by wind accretion or Roche-lobe
overflow) from a former AGB companion, which now most likely is a
white dwarf. In fact, many s-rich stars have been found to be
binaries \citep{McClure1980ApJ,McClure1990ApJ,North2000IAUS,
Barbuy2005A&A,Lucatello2006ApJ,Lucatello2009PASA}. A variety of
scenarios for explaining the abundance patterns of r+s stars have
been suggested (see, e.g., \citealt{Jonsell2006A&A}, and references
therein), but so far, none of them can coherently explain all
observational phenomena. Studies of additional r$+$s stars may shed
some lights on the general questions about the r- and s-processes,
such as the site or sites of the r-process and the relative
contribution of these two processes under metal-poor conditions.

To identify and study strongly r-process enhanced metal-poor stars,
i.e., r-II stars, the Hamburg/ESO R-process Enhanced Star survey
(HERES) has been carried out \citep{Christlieb2004A&A}. First, the
metal-poor candidates were selected in the digital spectra database
of the Hamburg/ESO objective-prism survey
\citep[HES;][]{Wisotzki2000A&A}. A detailed description of the
selection method and the method with which the metal-poor nature of
these candidates was confirmed based on their moderate-resolution
($\Delta\lambda\sim2$\,{\AA}) follow-up spectroscopy can be found in
\citet{Christlieb2008A&A}. During the course of HERES, ``snapshot''
spectra (i.e., spectra with $R=\lambda/\Delta\lambda = 20,000$ and a
typical signal-to-noise ratio of $S/N=50$) were obtained with the
Very Large Telescope (VLT) Unit Telescope 2 (UT2) and the
Ultraviolet-Visual Echelle Spectrograph (UVES) for several hundred
confirmed metal-poor stars. {\SampleStar}, the star studied here, is
one of them. It is a red giant star with a metallicity of
$\mathrm{[Fe/H]}\sim-2.4$, in which Eu and Ba are both enhanced.
Therefore, higher quality spectra of this star were obtained with
VLT/UVES (for details, see Sect. \ref{Sect:Observations}). Our
detailed abundance analysis is based on these spectra.

\section{Observations and data reduction}\label{Sect:Observations}

Astrometry and photometry of {\SampleStar} are listed in
Table~\ref{Tab:Photometry}; the photometry was taken from
\citet{Beers2007ApJS}. High-quality spectra of this object were
obtained during the night of 22 March 2005 with VLT-UT2 and UVES in
dichroic mode. The standard setting BLUE346+RED580 was used,
resulting in a spectral coverage of 3046--3863\,{\AA} in the blue
arm, and 4781--6809\,{\AA} in the red arm, with a gap between the
two CCD detectors causing a gap in wavelength coverage of
5757--5833\,{\AA}. Each spectrum has an exposure time of about 1hr,
and the total is 5hr. The slit width in both arms was set to
$0.8\arcsec$, so that a resolving power of $R=40\,000$ was achieved.
In the wavelength gap from $3850$\,{\AA} to $4795$\,{\AA} we used
the snapshot spectrum, which was obtained with VLT/UVES on 3 May
2003.

%
%
\begin{table}[htbp]
 \centering
 \caption{\label{Tab:Photometry} Photometry and astrometry of
   {\SampleStar}. The photometry was taken from \citet{Beers2007ApJS}.}
  \begin{tabular}{lr}
   \hline\hline
   R.A.(J2000.0) & $14\mathrm{h} 07\mathrm{m} 42.9\mathrm{s}$\\
   dec.(J2000.0) & $-08\degr 36\arcmin 14.3\arcsec$\\
   $V$          & $13.998\pm 0.003$\\
   $B-V$        & $ 0.772\pm 0.008$\\
   $V-R$        & $ 0.407\pm 0.005$\\
   $V-I$        & $ 0.827\pm 0.005$\\\hline
  \end{tabular}
\end{table}

The geocentric radial velocities of the individual spectra were
determined by fitting Gaussian profiles to $\sim 10$ moderately
strong, clean lines. Then the spectra were shifted to the rest
frame. The individual higher-resolution spectra were coadded in an
iterative procedure, in which pixels affected by cosmic-ray hits or
CCD defects were rejected by $\kappa\sigma$-clipping. The final
coadded spectrum was obtained by computing the weighted mean of the
individual spectra. In this coadded spectrum, the average
signal-to-noise ratio per pixel is $S/N\sim34$ from 3200 to
3800$\,\AA$ at Blue arm. The red-arm spectrum has $S/N>100$ per
pixel throughout the covered wavelength range from 4800 to
6700$\,\AA$. The snapshot spectrum has $S/N\sim50$ per pixel at
$4100\,\AA$.

The barycentric radial velocities of {\SampleStar} and the
observation epochs are listed in Table\,\ref{Tab:MJDRV}. The
difference of the radial velocity between the snapshot spectrum,
acquired at $\mathrm{MJD}=52762.213$, and the higher-resolution
spectra obtained 659 days later, is about 19\,km/s. This highly
significant radial velocity variation is a strong indication that
{\SampleStar} is a member of a binary system. Radial velocity
monitoring over several years will be needed to determine the period
and orbital parameters of the system.

\begin{table}[htbp]
 \centering
 \caption{\label{Tab:MJDRV} Barycentric radial velocities
 of {\SampleStar}}
  \begin{tabular}{lll}\hline\hline
    MJD    & $\mathrm{RV}$ & $\sigma$\\
    $\mathrm{[days]}$ & [km/s]                 & [km/s]  \\\hline
    $52762.213$ & $124.01$ & $0.55$\\
    $53451.178$ & $138.13$ & $0.17$\\
    $53451.214$ & $138.43$ & $0.53$\\
    $53451.251$ & $137.96$ & $0.35$\\
    $53451.288$ & $138.97$ & $0.71$\\
    $53451.325$ & $138.15$ & $0.45$\\\hline
  \end{tabular}
\end{table}

\section{Abundance analysis}\label{Sect:AbundanceAnalysis}

Our abundance analysis was carried out in local thermodynamic
equilibrium (LTE) conditions. Most of the Fe-peak and $\alpha$
abundances were determined by means of equivalent width
measurements. The analysis was restricted to lines with equivalent
widths more narrow than 100\,m{\AA} to avoid saturated lines and
potential fitting errors of Gaussian line profiles due to damping
wings that begin to appear at approximately this line strength. For
the other elements, the spectrum synthesis method was used,
employing the current version of the spectrum synthesis code
\citep[turbospectrum;\,][]{Alvarez1998A&A}. The snapshot spectrum
was used only for the abundances of crucial lines (e.g., Sr, CH, CN,
Eu), which are not present in the high-resolution UVES spectrum. The
abundance error is large for the snapshot spectrum for a given S/N
because of the lower resolution.

\subsection{Stellar parameters and model atmosphere}\label{Sect:stellarParameters}

An initial estimate of $T_{\mathrm{\scriptsize eff}}$ was determined
from broad-band optical and near-infrared colors, using the
calibrations of \citet{Alonso1996A&A} for $\mathrm{[Fe/H]}=-2.0$.
The transformation of 2MASS to TCS photometric system was the same
as was used in \citet{Sivarani2004A&A}. We adopted a reddening of
$E(B-V) = 0.037$ \citep{Schlegel1998ApJ}. The resulting effective
temperatures were used as an initial guess for the optimization
routine, as in \citet{Barklem2005A&A}. We additionally refined the
estimate using the line analysis procedure for the Fe~I and Ti~I
lines.

The optimization routine by \citet{Barklem2005A&A} assumes an
empirical relation for the microturbulence at various $\log\,g$. The
method does not independently estimate $T_{\mathrm{\scriptsize
eff}}$ and $\log\,g$. It finds a minimum chi-square solution by
fitting several weak metallic lines. However, the lines that are
sensitive to $\log\,g$ are strong neutral and ionized lines. Hence
there is a possibility of degeneracy between $T_{\mathrm{\scriptsize
eff}}$ and $\log\,g$. Therefore our additional refinement was to
independently estimate $T_{\mathrm{\scriptsize eff}}$, $\log\,g$,
and the microturbulence velocity using the Fe I and Ti I lines.
$T_{\mathrm{\scriptsize eff}}$ was estimated from the Fe I and Ti I
lines. $T_{\mathrm{\scriptsize eff}}$ was determined from the Fe I
lines by choosing a $T_{\mathrm{\scriptsize eff}}$ that does gives
no trend between the derived Fe abundance and its lower excitation
potential of the Fe I line. There are very few Ti lines, however,
therefore we used the Ti I lines only for a consistency check. The
values obtained with the different methods are listed in
Table~\ref{Tab:Parameters}. The surface gravity was derived from the
Fe~I/Fe~II ionization equilibrium, and a consistency check was made
using Ti~I/Ti~II ionization equilibrium. The microturbulence was
determined by requiring that the abundances derived from the Fe~I
lines be independent of the measured equivalent widths.

We employed OSMARCS model atmospheres (see
\citealt{Gustafsson2003IAUS} and references therein). Because
overabundances of C, N, and O may modify the temperature and density
structure of the atmosphere, we used a model atmosphere tailored for
{\SampleStar}, taking into account its enhancement in carbon,
nitrogen, and oxygen.

\begin{table}[ht]
 \caption{\label{Tab:Parameters} Stellar parameters of {\SampleStar}.}
  \begin{tabular}{lrrrrrl}
   \hline\hline
   Color/source/method & Value  & $T_{\mathrm{\scriptsize eff}}$ & $\log g$ &
   $\mathrm{[Fe/H]}$ & $v_{\mathrm{\scriptsize micro}}$ \\
   & \multicolumn{1}{c}{[mag]} & \multicolumn{1}{c}{[K]} & & &
     \multicolumn{1}{c}{[km/s]}\\\hline
   $B-V$                 & 0.772  & 4264 & \\
   $V-R$                 & 0.407  & 5321 & \\
   $V-I$                 & 0.827  & 5689 & \\
   $R-I$                 & 0.420  & 5305 & \\
   Barklem et al. (2005) &        & 5392 & 2.16 & $-2.27$ & 1.90 \\
   Fe~I \& Fe~II lines  &        & 5220 & 1.70 & $-2.40$ & 1.88 \\
   Adopted & & 5220 & 1.70 & -2.40 & 1.88 \\\hline\hline
  \end{tabular}
\end{table}

\subsection{Line selection and atomic data}\label{Sect:lineselection}

We used the CH nd CN molecular line list compiled by
\citet{Plez2005IAUS}. The NH and C$_{2}$ molecular line lists were
taken from the Kurucz
database\footnote{http://kurucz.harvard.edu/linelists/linesmol/}.

The line data for Pb were taken from \citet{VanEck2003A&A}.
\citet{McWilliam1995AJ} pointed out that hyperfine splitting (HFS
hereafter) has the effect of desaturating strong lines. Hence it is
very important to perform HFS for strong lines. We included HFS and
isotopic fractions as given in \citet{VanEck2003A&A}. We detected
seven Ba~II lines. We adopted the HFS provided by
\citet{McWilliam1998AJ}. The Eu line list is the same as in
\citet{Mucciarelli2008A&A}, who adopted the values from
\citet{Lawler2001ApJ}. We also checked the difference between the
HFS provided by \citet{Kurucz1993sssp.book} and
\citet{Lawler2001ApJ}. The derived abundances agreed well. The $gf$
values of the La lines were taken from \citet{Lawler2001ApJL}, and
the HFS provided by \citet{Ivans2006ApJ} were also considered.
According to \citet{McWilliam1995AJ}, HFS is important for any La
lines with equivalent widths (EW) greater than
$log_{10}(EW/\lambda)>-5.6$. Hence many of the lines used are
probably affected by HFS. The atomic data for the other lines comes
from the Vienna Atomic Line Database (VALD). The HFS of Pr and Yb
lines  provided by \citet{Ivarsson2001PhyS}, \citet{Sneden2009ApJS},
Yb, \citet{Biemont1998JPhB}, and \citet{Sneden2009ApJS},
respectively were also adopted. The Yb\,II abundances are based on
the 3694.192$\,\AA$ line. We did not use the Yb\,II 3289.367$\,\AA$
line, because it shows blending from other atomic lines (e.g. V\,II
and Fe\,II). We derived the Ce, Nd, and Y abundances from weak
narrow lines. They probably have no significantly resolved structure
because of HFS at the observed spectral resolution.

The selected lines are listed in
Table~\ref{Tab:EWlinelist},~\ref{Tab:synthlinelist} (Online
material), along with the transition information and references to
the adopted $gf-$values.

\section{Abundance results}\label{Sect:abundanceresults}

We derived abundances for 39 elements. When elemental abundances for
a species were derived from multiple lines, we adopted the the error
of the mean (i.e.,
$\sigma\left(\log\left(\epsilon\right)\right)/\sqrt{(N)}$) as the
uncertainty of the abundance measurement of the species.  For the
spectrum synthesis measurement, we estimated an the uncertainty
based on the Cayrel formula \citep{Cayrel1988IAUS}, yielding
$0.02$--$0.12$\,dex at the $S/N $ of the red and blue spectrum,
respectively, and 0.2\,dex for lines detected only in the snapshot
spectrum. The total errors in [X/Fe] for each element are about
$0.02$--$0.30$\,dex, taking into account uncertainties of 150\,K in
$T_{\mathrm{\scriptsize eff}}$, and 0.5\,dex in $\log g$, which were
estimated using a weighted mean of the various estimates listed in
table~\ref{Tab:Parameters}. In Table~\ref{Tab:AbundanceSummary} we
list the mean abundances ($\eps{}$), the mean
errors($\sigma_{\eps{}}$), the number of lines used to determine the
mean abundances, and the abundances relative to iron ([X/Fe]). We
adopted the solar abundance of \citet{Grevesse1998SSR}.

\begin{table} 
 \caption{\label{Tab:AbundanceSummary} Summary of the abundances of
   {\SampleStar}.}
 \centering
 \begin{tabular}{rlrrrrl}\hline\hline
   $Z$ & Species  & $N_{\mathrm{\scriptsize lines}}$ & $\log\epsilon$ &
   $\sigma_{\eps{}}$ & $\mathrm{[X/Fe]}$ & Notes \\\hline
   3  & Li I  &  2 & $0.90$   & 0.20 & $     $ & Synth\\
   4  & Be II &  2 & $<-2.80$ & 1.00 & $-1.72$ & Synth\\
   6  & CH    &  1 & $7.96$   & 0.10 & $ 1.97$ & Synth\\
   6  & C$_2$ &  1 & $7.96$   & 0.12 & $ 1.97$ & Synth\\
   7  & NH    &  1 & $6.80$   & 0.15 & $ 1.34$ & Synth\\
   7  & CN    &  1 & $6.80$   & 0.15 & $ 1.34$ & Synth\\
   8  & OH    &  2 & $7.60$   & 0.20 & $ 1.27$ & Synth\\
   11 & Na I  &  2 & $4.62$ & 0.15 & $ 0.73$ & Synth\\
   12 & Mg I  & 10 & $5.55$ & 0.12 & $ 0.41$ & EW\&synth\\
   13 & Al I  &  2 & $3.07$ & 0.10 & $-0.99$ & Synth\\
   14 & Si I  &  3 & $4.55$   & 0.14 & $-0.58$ & EW\& synth\\
   20 & Ca I  & 11 & $4.26$ & 0.10 & $ 0.35$ & EW\\
   21 & Sc II & 3  & $1.15$   & 0.10 & $ 0.48$ & EW\\
   22 & Ti I  & 9  & $2.85$   & 0.10 & $ 0.32$ & EW\\
   22 & Ti II & 9  & $2.99$   & 0.10 & $ 0.32$ & EW\\
   23 & V II  & 3  & $1.56$   & 0.24 & $-0.03$ & Synth\\
   24 & Cr I  & 18 & $3.16$   & 0.10 & $-0.09$ & EW\&synth\\
   25 & Mn I  & 3  & $2.61$   & 0.25 & $-0.49$ & Synth\\
   25 & Mn II & 8  & $2.61$   & 0.12 & $-0.49$ & Synth\\
   26 & Fe I  & 74 & $5.13$   & 0.10 &         & EW\\
   26 & Fe II &  6 & $5.13$   & 0.10 &         & EW\\
   27 & Co I  &  2 & $2.36$   & 0.17 & $-0.14$ & Synth\\   
   28 & Ni I  &  5 & $4.00$   & 0.12 & $ 0.18$ & EW\&synth\\    
   29 & Cu I  &  2 & $0.70$   & 0.11 & $-1.17$ & Synth\\
   30 & Zn I  &  2 & $2.50$   & 0.02 & $ 0.25$ & EW\\
   38 & Sr II &  2 & $0.32$   & 0.04 & $-0.18$ & Synth\\
   39 & Y II  & 13 & $0.10$   & 0.10 & $ 0.30$ & Synth\\
   40 & Zr II & 16 & $0.97$   & 0.11 & $ 0.80$ & Synth\\
   41 & Nb II & 2  & $0$   & 0.3  & $ 0.98$ & Synth\\
   56 & Ba II & 7  & $1.73$   & 0.17 & $ 1.95$ & Synth\\
   57 & La II & 13 & $0.26$   & 0.16 & $ 1.47$ & Synth\\
   58 & Ce II & 15 & $0.15$   & 0.12 & $ 0.95$ & Synth\\
   59 & Pr II &  8 & $-0.19$  & 0.13 & $ 1.44$ & Synth\\
   60 & Nd II & 17 & $0.70$   & 0.12 & $ 1.63$ & Synth\\
   62 & Sm II &  5 & $-0.31$  & 0.20 & $ 1.13$ & Synth\\
   63 & Eu II &  2 & $-0.33$  & 0.20 & $ 1.54$ & Synth\\
   64 & Gd II &  5 & $-0.19$  & 0.20 & $ 1.12$ & Synth\\
   65 & Tb II &  1 & $-0.98$  & 0.20 & $ 1.08$ & Synth\\
   66 & Dy II &  7 & $-0.19$  & 0.20 & $ 1.07$ & Synth\\
   68 & Er II &  1 & $-0.23$  & 0.25 & $ 1.22$ & Synth\\
   70 & Yb II &  2 & $0.40$   & 0.20 & $ 1.86$ & Synth\\
   71 & Lu II &  3 & $-1.07$  & 0.20 & $ 1.22$ & Synth\\
   72 & Hf II & 12 & $0.09$   & 0.11 & $ 1.76$ & Synth\\
   82 & Pb I  & 2  & $1.96$   & 0.20 & $ 2.30$ & Synth\\\hline
\end{tabular}
\end{table}

\section{Abundance pattern of HE~1405$-$0822}\label{Sect:Observedpattern}

\subsection{n-capture elements}\label{Sect:n-capture}

\begin{figure*}[ht]
  \centering
  \includegraphics[bb=0 142 564 720,width=12cm,clip]{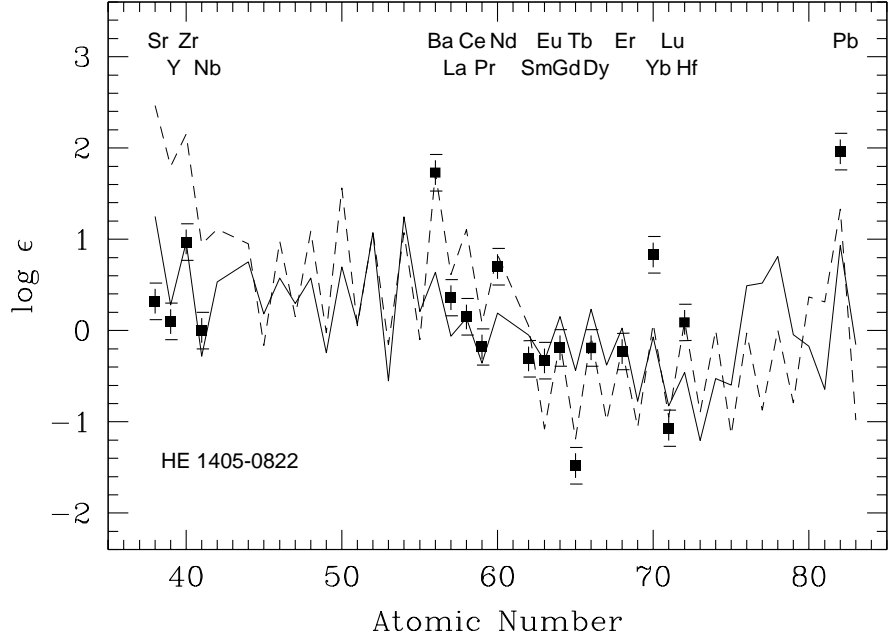}
  \vspace{-2.5cm}
  \caption{\label{Fig:Solarrs} Observed neutron-capture element
    abundances of {\SampleStar} (full squares) compared with the
    scaled solar s- and r-process abundance patterns (dashed and solid
    lines, respectively). The solar s-process pattern was normalized to
    Ba, the solar r-process pattern was normalized to Eu. }
\end{figure*}

{\SampleStar} is a carbon-enhanced metal-poor \citep[CEMP, for the
definition see][]{Beers2005ARA&A} r$+$s star, whose n-capture
elements exhibit a high overabundance relative to Fe and the
abundance ratios in the Sun. The only exception is Sr, which is
underabundant ($\mathrm{[Sr/Fe]}=-0.18$). For this star, the ratio
of [La/Eu], which is a good indicator of the s- and r-process
contribution in stars, is $-0.07$. Indeed, 75\,\% of the solar La is
synthesized by the s-process, while about 97\,\% of the solar Eu
originates from the r-process \citep{Burris2000ApJ}. Judging from
its Eu and La abundance ratios, $\mathrm{[Eu/Fe]}=1.54$ and
$\mathrm{[La/Eu]}=-0.07$, {\SampleStar} has experienced a major
r-process-enrichment event.

Figure~\ref{Fig:Solarrs} shows neither the scaled solar r-pattern
nor the scaled solar s-pattern agrees with the neutron-capture
element abundance pattern of {\SampleStar}. The two ratios
[hs/ls]\footnote{Here we adopted the average of Ba and La
  as 'hs', representing the second s-process peak, and the average of Y
  and Zr as 'ls', representing the first s-process peak.} and [Pb/hs]
are good indicators of the $^{13}$C-pocket efficiency in AGB stars,
which are independent of the efficiency of the third dredge-up (TDU
hereafter) event as well as of the dilutions of the s-process
synthetic material both in the AGB envelope and in the secondary of
the binary system. The abundance ratios [hs/ls] and [Pb/hs] of
{\SampleStar} are $1.16$ and $0.59$. The $^{13}$C-pocket in the AGB
star is clearly highly efficient. This is responsible for the
significant enhancement of heavy s-process elements such as Ba, La,
and Pb. But, this star does not belong to the so-called lead stars
\citep[$\mathrm{[Pb/hs]}\ge 1.0$, see
][]{Gallino1998ApJ,Goriely2000A&A,Goriely2001A&A} such as HD~187861,
HD~224959, or HD~196944 \citep{VanEck2001Natur}. This means that the
efficiency of the $^{13}$C-pocket is not high enough to provide
sufficient neutrons for a large number of Pb nuclei, which are close
to the termination point of the s-process path. In addition, because
the $^{22}$Ne neutron source mainly contributes to the first
s-process peak, the negative [Sr/Fe] ratio in {\SampleStar}
indicates that its former AGB companion probably had a relatively
low mass of $M<3$\,M$_\odot$ \citep{Bisterzo2010MNRAS} or
$M<4$\,M$_\odot$ \citep{Karakas2007PASA}, where the $^{22}$Ne
neutron source works only marginally for the s-process during the
thermal pulses during the AGB phase.

\subsection{Possible formation mechanism}\label{Sect:formationmechanism}

We compared the observed abundance distribution of {\SampleStar}
with the results of theoretical r- and s-process nucleosynthesis
calculations. We used the parametric model for metal-poor stars
presented by \citet{Zhang2006ApJ} and developed by
\citet{Cui2007ApJ} and \citet{Cui2010ApJ}. In the model, we
calculated the envelope abundance $N_{i}$ of the $i$th element as
follows:
\begin{equation}
N_{i}(Z)=C_{s}N_{i,\ s}+C_rN_{i,\ r}10^{[Fe/H]},
\end{equation}
where $Z$ is the metallicity of the star, $N_{i,\ s}$ and $N_{i,\
r}$ are the abundance of the $i$th element produced by the s- and
r-process (per Si $=10^6$ at $Z=Z_\odot$) and $C_s$ and $C_r$ are
the component coefficients representing the contributions of the s-
and the r-process. We assumed that {\SampleStar} formed from a gas
cloud that was enriched by an r-process nucleosynthesis event. Thus,
the scaled solar r-element abundance was adopted as the initial
abundance $N_{i,\ r}$. $N_{i,\ s}$ was calculated from the
parametric model by means of an extensive reaction network described
earlier \citep{Liang2000A&A}. Because in s-rich stars the weak
r-process only marginally contributes to the production of light
n-capture elements such as Sr, Y, and Zr. \citep{Liang2012PASP}
compared with the s-process, we ignored the weak r-process
contribution in this work.

\begin{figure*}[htbp]
  \centering
  \includegraphics[bb=0 142 564 720,width=12cm,clip]{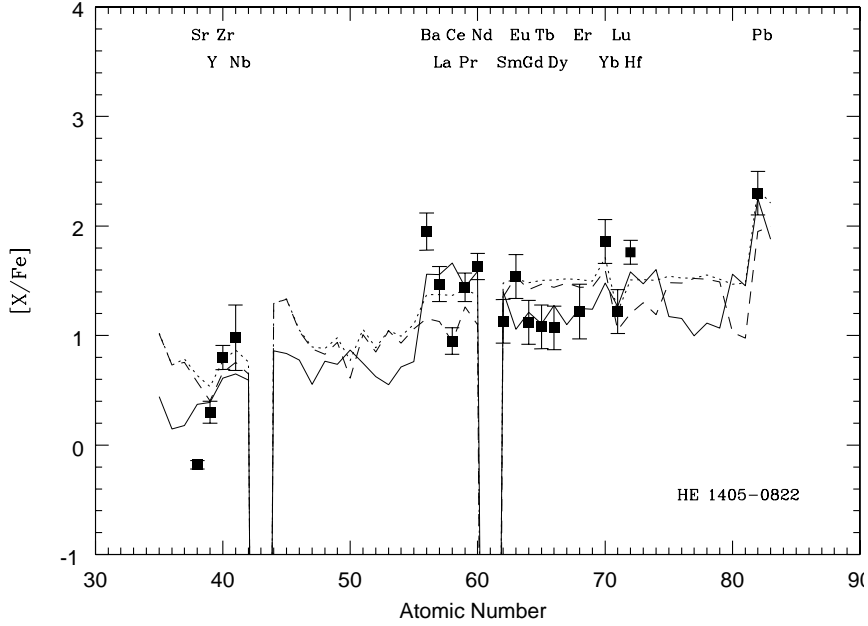}
  \vspace{-2.5cm}
  \caption{\label{Fig:xfe} Abundance pattern of {\SampleStar} (full
    squares) compared with theoretical predictions. The solid line
    represents our results based on a parametric method
    \citep{Zhang2006ApJ,Cui2010ApJ}. The other two lines represent the results of two AGB models
    \citep{Bisterzo2010MNRAS} with $\mathrm{[Fe/H]}=-2.6$, i.e. $2.0\,M_\odot$,
    ST/6, $\mathrm{[r/Fe]}=0.9$, $dil\,=1.4$ (dashed line), and
    $1.4\,M_\odot$, ST/6, $\mathrm{[r/Fe]}=1.0$, $dil\,=1.1$ (dotted
    line).}
\end{figure*}

In Figure~\ref{Fig:xfe} ST refers to the standard case of the
$^{13}$C-pocket including about $3.0\times10^{-6}\,M_\odot$ of
$^{13}$C and $9.0\times10^{-8}\,M_\odot$ of $^{14}$N adopted by
\citet{Gallino1998ApJ}, which can reproduce the s-process main
component of the solar system using an AGB model with $Z_{\odot}/2$.
The parameter $dil$ is the dilution factor; i.e., the degree of
dilution of the AGB material after accretion by the lower-mass
companion in a binary system. This low-mass companion is the star
that we observe today. [r/Fe] is the degree of initial r-process
enrichment in the gas cloud from which the binary formed.

From Figure~\ref{Fig:xfe} we can see that most of the 19 observed
heavy neutron-capture elements agree with our theoretical
predictions within the measurement uncertainties of the abundances.
The s-process ratios observed in {\SampleStar},
$\mathrm{[Pb/hs]}=0.59$ and $\mathrm{[hs/ls]}=1.16$, also agree with
the predictions of the parametric method within the errors; these
predictions are $\mathrm{[Pb/hs]}=0.69$ and $\mathrm{[hs/ls]}=1.21$.
This strongly supports the reliability of our obtained
nucleosynthesis parameters, i.e., the neutron exposure per thermal
pulse $\Delta\tau=0.69$\,mbarn$^{-1}$, the overlap factor $r=0.49$,
the component coefficient of the s-process $C_{s}=0.00095$, and the
component coefficient of the r-process $C_{r}=8.7$, where the
overlap factor $r$ is the fraction of material in the He intershell
of an AGB star that still has to experience subsequent neutron
exposures. $C_{r}$ and $C_{s}$ are the component coefficients that
correspond to the s- and r-process contributions.

The mean neutron exposure for {\SampleStar} is
$\tau_0=1.81(T_9/0.348)^{1/2}$, where $T_9=0.1$ (in units of
$10^9\,$K). \citet{Kappeler1989RPPh} found
$\tau_0=0.30(T_9/0.348)^{1/2}$ when they fit the solar main
component. Based on the primary nature of the $^{13}$C source,
\citet{Gallino1998ApJ} found a maximum neutron exposure of
$0.40$--$0.45$\,mbarn$^{-1}$ with their standard AGB model, which
can reproduce the solar s-process distribution well. Furthermore,
they also pointed out that the average s-process efficiency will
indeed increase toward lower metallicity, which is mainly due to the
decreasing iron abundance, and therefore higher neutron-to-seed
ratio. The values of $\Delta\tau$ and $\tau_0$ for {\SampleStar} are
significantly higher than those for the solar system. This can
naturally explain why the enrichment of the s-process material in
{\SampleStar} is significantly stronger than in the solar system.

The overlap factor for {\SampleStar}, $r=0.49$, lies in the range of
$r \sim 0.4$--$0.7$ found by \citet{Gallino1998ApJ}, using their
standard low-mass AGB model at solar metallicity. Using an s-process
parametric model without adopting any specific stellar model,
\citet{Aoki2001ApJ} reported a neutron exposure per pulse of about
$0.7$--$0.8$\,mbarn$^{-1}$, and a small overlap factor of $\sim0.1$
for two carbon-rich metal-poor r$+$s stars, LP\,625$-$44 and
LP\,706$-$7, with $\mathrm{[Fe/H]}=-2.7$. That is, these two stars
have similar values of the neutron exposure per pulse as
{\SampleStar}, but significantly lower values of the overlap factor
than {\SampleStar}. \citet{Aoki2001ApJ} proposed a new mechanism for
the s-process, a single neutron-exposure event. They found that
during the first neutron exposure almost all elements except Pb can
be produced in their parametric model. Even the Pb abundance can be
reproduced after about three recurring neutron exposures, which
corresponds to a small overlap factor of $r\lesssim0.2$. In
conclusion, a single neutron-exposure event of the s-process for
{\SampleStar} can be excluded.

Because {\SampleStar} is in its red giant evolution phase, the
s-elements cannot be synthesized by themselves. Instead, the
s-elements were probably synthesized by its former AGB companion in
a binary system and were then transferred to its surface by a
stellar wind. The component coefficient of the s-process,
$C_s=0.00095$, is very small. Therefore, the binary system probably
had a long orbital period, which results in a small amount of
material that is accreted.

Regarding the origin of the r-process component of the abundance
pattern of {\SampleStar}, the accretion-induced collapse mechanism
\citep{Qian2003ApJ,Cohen2003ApJ} can be excluded for a long orbital
period, because it makes the opposite mass accretion difficult to
imagine, i.e. the white dwarfs (the remnant of its AGB companion)
accreting material from the secondary star observed now.
Interestingly, our parametric calculation fitted almost all
r-process element abundances such as Gd (r-process fraction in the
solar system $82$\,\%; \citealt{Burris2000ApJ}), Tb ($94$\,\%), Dy
($88$\,\%), Er ($84$\,\%), and Lu ($79$\,\%), but underestimated the
Eu ($97$\,\%) and Yb (68\%) abundance. The predictions of low-mass
AGB model \citep{Bisterzo2010MNRAS} with two different initial
masses (see dashed and dotted lines) where the r-process
pre-enrichment scenario (formed from a cloud which have been
polluted by SNe of type II) were adopted are also plotted in
Figure~\ref{Fig:xfe} for comparison. For the low-mass AGB model
calculation, the r-element pre-enriched mechanism was adopted, that
is, we adopted solar r-element abundances, which scaled to Eu of
{\SampleStar} as the initial model values. From Figure~\ref{Fig:xfe}
we can see that most r-process elements of {\SampleStar} are
overestimated except for Eu and Yb.

We adopted the solar r-process pattern to calculate the main
r-process contribution in all model calculations discussed above.
However, the abundance pattern of the r-process contribution in
{\SampleStar} is inconsistent with the scaled solar pattern,
therefore it is also inconsistent with the universal pattern
observed in r-II stars \citep[and references
therein]{Sneden2008ARA&A}. This is in stark contrast to what was
found for instance in the r$+$s star HE~0338$-$3945
\citep{Cui2010ApJ}. Compared with the solar r-process pattern, the
incongruously high Eu abundance of {\SampleStar} relative to other
second-r-process-peak elements such as Gd and Tb may be caused by
observational uncertainties. If this is not the case, a more complex
origin for the r-process is implied, especially for the r+s stars.

\citet{Lugaro2012ApJ} studied many CEMP-s and CEMP-r$+$s stars with
their detailed AGB evolution models. They found that the r-process
pre-enrichment scenario mainly have three problems for explaining
the formation of CEMP-r$+$s stars. (1), They were unable to
reproduce the linear correlation observed between Ba and Eu
enrichments in the currently known sample of CEMP-r$+$s stars,
because the initial [r/Fe] value does not affect the final [Ba/Fe]
value in an AGB model. In other words, in the pre-enrichment
scenario the two independent nucleosynthesis processes (i.e., r- and
s-process) who do not affect each other cannot reproduce the
Ba-Eu-enrichments correlation in CEMP-r$+$s stars. (2), It is
difficult to explain the smaller number of r-II stars (about 10)
compared with CEMP-r$+$s stars (about 30), because in this scenario
CEMP-r$+$s stars should be formed from r-II stars. (3), Because of
the similar r-elements origin in this scenario for r-II and
CEMP-r$+$s stars, the different metallicity distribution of r-II
stars at [Fe/H]$\simeq-2.8$ and CEMP-r$+$s stars at
[Fe/H]$\simeq-2.5$ is difficult to explain. Thus, they argued that
the r-process seen in r$+$s stars is different from that observed in
r-II stars, and that even the results of s-process nucleosynthesis
seen in r$+$s stars is different from that seen in CEMP-s stars
because of their typically higher Ba abundances. Because the
pre-enrichment scenario seems difficult to determine the origin of
the r-elements in r$+$s stars, some other forms of nucleosynthesis
must be responsible. To explain this, \citet{Lugaro2012ApJ} assumed
an ``s/r" neutron-capture process, which they described as a single
process with features that are similar to or an addition of the s-
and r-process. If this is true, it should produce the positive
correlations between Ba and Eu abundances in r$+$s stars and
possibly r-process patterns different from that of the Sun. However,
this hypothesis still needs theoretical confirmation. These authors
also considered a model involving a stable triple stellar system
\citep[for details see][]{Jonsell2006A&A}, despite the unstability
problem of the dynamics and the low occurrence likelihood. In the
triple system, the primary exploded as an SNe of type II (hereafter
SN~II) and produced r-elements, and the other companion polluted the
observed star during its AGB phase with s-rich material. Such
scenarios may offer a solution for the r-process origin of
{\SampleStar}. We also cannot exclude the scenario in which the
binary system formed from a gas cloud that was enriched with
r-process material. But this would imply that the enrichment event
would have resulted in an abundance pattern that at least in some
cases is different from the r-process pattern seen in the Sun and
r-II stars.

We did not include NLTE corrections for any of the neutron-capture
elements, but many lines used in the analysis may need NLTE
corrections. For Sr\,II line 4077\,{\AA}, based on
\citet{Bergemann2012A&A} and \citet{Belyakova1997AZh}, we found that
the NLTE corrections for {\SampleStar} is about -0.01\,dex. However,
\citet{Mashonkina2000A&A} reported positive NLTE corrections for the
Eu~II resonance line at 4129\,{\AA} and the subordinate line at
6645\,{\AA}, which was confirmed by \citet{Asplund2005ARA&A}. Based
on \citet{Mashonkina2012A&A}, the NLTE corrections for the Eu II
lines 4129 and 4205\,{\AA} of {\SampleStar} are probably about
0.1\,dex . The NLTE corrections for Pb\,I is very high, 0.5-0.6 dex
\citep{Mashonkina2012A&A}.

\subsection{Elements up to the iron peak}\label{Sect:lightelements}

Like many other r$+$s stars, {\SampleStar} also exhibits strong
enhancements of carbon, nitrogen, and oxygen. In this star, sodium
and magnesium are also enhanced. Because we cannot calculate the
abundances of the elements up to the iron peak with our parametric
method for the s-process, we only compared the observed abundances
with the AGB model yields of \citet{Bisterzo2010MNRAS} and the
observed abundances of CS 22892$-$052 \citep{Sneden2003ApJ} which
normalized to the iron abundance of {\SampleStar}.

\begin{figure*}[ht]
  \centering
  \includegraphics[bb=0 142 564 720,width=12cm,clip]{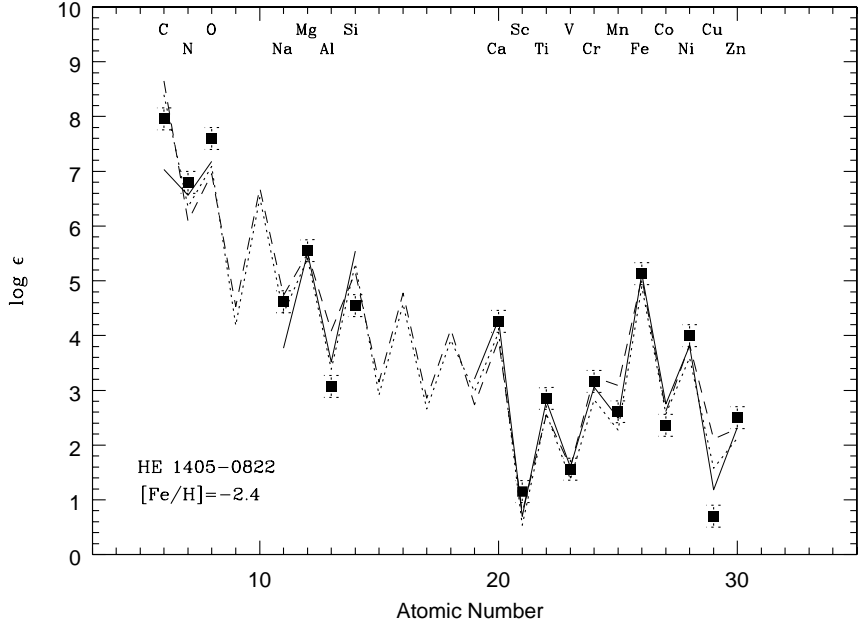}
  \vspace{-2.5cm}
  \caption{\label{Fig:cnom} Abundance pattern of {\SampleStar} (full
    squares) compared with theoretical predictions. The solid line
    represents the observed distribution of CS 22892$-$052 normalized to
    iron. The other two lines represent the results of AGB models
    \citep{Bisterzo2010MNRAS} with $\mathrm{[Fe/H]}=-2.6$, i.e. $2.0\,M_\odot$,
    ST/6, $\mathrm{[r/Fe]}=0.9$, $dil\,=1.4$ (dashed line), and
    $1.4\,M_\odot$, ST/6, $\mathrm{[r/Fe]}=1.0$, $dil\,=1.1$ (dotted
    line).}
\end{figure*}

CS 22892$-$052 is an r-II star. Its heavy neutron-capture-element
($Z\ge56$) abundances agree well with a scaled solar r-process
abundance pattern. Generally, SN~II are thought to be responsible
for the production of heavy r-process pattern, because of the low
metallicities ($\mathrm{[Fe/H]}\sim3.0$) of the observed r-II stars,
which indicates that the r-process sites must be short-lived and
have evolved rapidly, so that the interstellar medium (ISM) could be
enriched in r-elements prior to the formation of the r-II stars.
Indeed, the enhanced $\alpha$-elements such as O, Mg, and Si in CS
22892$-$052 are thought to be generated by the pollution by SN~II.
From Figure~\ref{Fig:cnom} we can see that the observed abundances
of the elements from Ca to Zn in {\SampleStar} agree well with the
scaled ones of CS 22892$-$052 and also with the predictions of
\citet{Bisterzo2010MNRAS}. This means that the abundances of these
elements did not change during the evolution of the binary system,
but remained at the values of ISM at the time when and location
where {\SampleStar} formed.

From Figure~\ref{Fig:cnom} we can see that neither the AGB model
predictions nor the scaled abundances of CS 22892$-$052 fit the
carbon, nitrogen, and oxygen abundances of {\SampleStar} well. For
the C and N abundances of CS 22892$-$052 and {\SampleStar}, both CH
and CN features were used. The AGB models of
\citet{Bisterzo2010MNRAS} often overestimate the carbon and oxygen
abundance and underestimate the nitrogen abundance compared with
stars that enriched in neutron-capture elements
\citep{Bisterzo2011MNRAS}. This may be due to the model itself, for
instance for the incorrect yields of C, N, and O, or the uncertainty
of the observed abundances. Due to the strong temperature
sensitivity of CH and CN molecular lines, the uncertainty of the
derived molecular-based C and N abundances is large at low
metallicity. 3D corrections for C and N can reach about $-0.5$ to
$-0.3$\,dex at low metallicity
\citep{Asplund2001A&A,Asplund2004MmSAI}. However, since the
magnitudes of the 3D corrections of these two elements are roughly
the same, the C/N ratio is probably not strongly affected. We recall
that the 3D corrections are themselves highly model-dependent and
highly uncertain at present. In {\SampleStar}, this ratio is
$\mathrm{C/N}=14$, which strongly suggests that hot-bottom burning
(HBB) did not occur in the former AGB companion, because if HBB had
occured, the observed C/N ratio would be a constant 1/15
\citep{McSaveney2007MNRAS} or 1/10 \citep{Herwig2004ApJS}. Using
detailed evolution models of AGB stars, \citet{Karakas2007PASA}
showed that the lowest mass limit is between $2.5$ and
$3$\,M$_\odot$, where HBB could set in around
$\mathrm{[Fe/H]}=-2.3$. This is consistent with the low-mass
estimate for the former AGB companion of {\SampleStar} presented in
Sect.~\ref{Sect:n-capture} and Sect.~\ref{Sect:formationmechanism}.

Non-LTE effects of UV OH molecular line formation may be strong, but
NLTE corrections are not available. 3D effects and missing line
opacities are thought be an important uncertainty in deriving an
abundance from UV-OH lines. \citet{Garcia2006A&A}, however, found a
good agreement between the oxygen abundance derived from [O I] and
UV-OH lines. According to \citet{Herwig2004ApJS} and
\citet{Sivarani2006A&A}, low-mass AGB also produce some oxygen,
which does not change the initial oxygen abundance significantly at
higher metallicities. Furthermore, it could increase the oxygen
abundances for low metallicity stars. But we can from
figure~\ref{Fig:cnom} see that the high oxygen abundance of
{\SampleStar} can be reached neither by CS 22892$-$052 nor by AGB
predictions \citep{Bisterzo2011MNRAS}. OH lines were used here for
the oxygen abundances of {\SampleStar}, while Sneden et al. (2003)
used the [O\,I] 6300 feature for CS 22892$-$052. High oxygen
abundance is also seen in many r+s stars \citep{Masseron2010A&A}.
For the dilution effect, \citet{Cui2010ApJ} pointed out that the
oxygen abundance contributed by the low-mass AGB stars could only
reach $\mathrm{[O/Fe]}\sim0.8$, for $\mathrm{[C/Fe]}\sim2.0$ based
on the AGB model of \citet{Karakas2007PASA}. The possible origin for
most of the oxygen in {\SampleStar} is probably an SN~II, a similar
origin as for CS 22892$-$052.

We also found sodium and magnesium to be enhanced in {\SampleStar},
i.e. $\mathrm{[Na/Fe]}=0.73$ and $\mathrm{[Mg/Fe]}=0.41$. For Na and
Mg the observations and AGB model predictions in
Figure~\ref{Fig:cnom} match well. The scaled Mg abundance of CS
22892$-$052 agrees well with the Mg abundance of {\SampleStar}, but
this is not the case for Na.  Again, we need to consider NLTE
corrections before we can draw any conclusions. The Na D lines
($\lambda5889$ and $\lambda5895$) are significantly affected by
NLTE, which are used in the present work as well for CS 22898-052.
For Mg abundances, there are at least three common Mg\,I features
used for both {\SampleStar} and CS 22892$-$052. In metal-poor stars,
typical corrections for Na are about $-0.3$\,dex
\citep{Andrievsky2007A&A}, and for Mg about $+0.2$\,dex
\citep{Aoki2007ApJ,Andrievsky2010A&A}. Since the stellar parameters
are similar for CS 22892$-$052 and {\SampleStar}, the NLTE
corrections of Na and Mg are probably also similar for these two
stars.

Even taking into account typical NLTE corrections, the enhancement
of Na and Mg in {\SampleStar} could still be explained by mass
transfer from a lower-mass AGB companion, where the primary
$^{22}$Ne mainly operated as a neutron poison in the
$^{13}$C-pocket. In low-mass AGB stars of low metallicity
($\mathrm{[Fe/H]}<-1.0$), a primary production of Na and Mg can be
generated by the reactions $^{22}$Ne(n,$\gamma$)$^{23}$Na, and then
$^{23}$Na(n,$\gamma$)$^{24}$Mg, $^{22}$Ne($\alpha$,n)$^{25}$Mg,
$^{22}$Ne($\alpha$,$\gamma$)$^{26}$Mg
\citep{Mowlavi1999A&A,Gallino2006isna.confE}. The primary production
of $^{22}$Ne increases with the initial mass of the AGB star at very
low metallicity \citep{Bisterzo2006MmSAI}. The light enrichment of
Mg in {\SampleStar} also supports a low-mass AGB companion, since
there is not enough $^{22}$Ne to feed a higher Mg abundance. In
addition, because $\mathrm{[Mg/Fe]}=0.41$ of {\SampleStar} is very
similar to the value seen in other field stars, a common origin of
Mg, that is pre-enriched by SN~II
\citep{Gehren2006A&A,Andrievsky2010A&A}, should not be excluded.

In {\SampleStar}, Al is strongly underabundant;
$\mathrm{[Al/Fe]}=-0.99$. The NLTE correction is expected to be
about $0.15$\,dex \citep{Andrievsky2008A&A}. If applied, the Al
abundance would match the theoretical prediction of an AGB model
with $M=1.4\,\mathrm{M}_\odot$ \citep{Bisterzo2010MNRAS}. The low
abundance of Al also supports our assumption, i.e. a low-mass former
AGB companion of {\SampleStar}, which contributed little primary Al
\citep{Karakas2007PASA}. This means that the observed Al probably
also comes from the ISM at the time and place where {\SampleStar}
formed.

The differences between the Si abundances of {\SampleStar}, the
scaled abundances of CS 22892$-$052, and the AGB model predictions
(see Figure~\ref{Fig:cnom}) cannot be explained by the small
negative NLTE correction of about $-0.05$\,dex determined by
\citet{Shi2009A&A,Shi2011A&A}. A possible explanation are different
masses and yields of the SN~II that pre-enriched {\SampleStar} and
CS 22892$-$052. \citet{Preston2006AJ} discussed the systematic
effects in the Si abundances caused by the use of different lines.
Si abundances of {\SampleStar} are derived from Si I 3905.523 and
4102.936\,\AA\,lines. \citet{Sneden2003ApJ} also used the Si\,I
4102.94\,\AA\,line for CS\,22892-052.

\section{Conclusions}\label{Sect:DiscussionConclusions}

We have analyzed high-quality VLT/UVES spectra of {\SampleStar} and
derived accurate abundances for 39 elements, including 19
neutron-capture elements. {\SampleStar} shows strong enhancement in
both the r-process and s-process elements (e.g.,
$\mathrm{[Eu/Fe]}=1.54$, $\mathrm{[Ba/Fe]}=1.95$, and
$\mathrm{[Pb/Fe]}=2.3$), therefore we confirm that it is an r$+$s
star.

We discussed several scenarios for the origin of the abundance
pattern of {\SampleStar}, taking into account the possible influence
of NLTE and 3D corrections on the interpretation of our results.

Because {\SampleStar} is in its red giant evolutionary phase, it
cannot produced its strong enhancement of C, N, and s-elements by
itself. Instead, it is very likely that these enhancements were
produced in a formerly more massive companion during its AGB phase
and were transferred to the surface of the star that we observe
today. The binarity of {\SampleStar} is confirmed by the fact that
the star shows significant radial velocity variations. Combining the
enriched s-process material and significant radial velocity
variations of {\SampleStar}, its pre-AGB companion probably is a
white dwarf now. However, excess UV-flux measurements are also
needed to confirm this. In addition, we also need long-term radial
velocity monitoring to confirm its binary nature, and to determine
its orbital period and parameters.

Neither the scaled solar s-process pattern nor the scaled solar
r-process pattern match the observed abundance pattern of
{\SampleStar} well. We compared the abundance pattern with
predictions of our parametric method \citep{Zhang2006ApJ,Cui2010ApJ}
and two AGB model yields of \citet{Bisterzo2010MNRAS}. In both
cases, the Pb to heavy s-process element ratio of this star
($\mathrm{[Pb/hs]}=0.59$) and the heavy-to-light s-process element
ratio ($\mathrm{[hs/ls]}=1.16$) can be reproduced by the models.
This strongly supports the reliability of the s-process calculations
considered in this work. The parameter fits of the models yield the
result that the AGB companion probably is a star with relatively low
initial mass, about $\le 2\,\mathrm{M}_\odot$. This is supported by
the low Sr/Fe ratio of $\mathrm{[Sr/Fe]}=-0.99$, which means that
the $^{22}\mathrm{Ne}$ neutron source had only a mild influence on
the s-process that occurred in the former AGB companion.

Unlike in HE~0338$-$3945, we cannot reproduce the r-process pattern
of {\SampleStar} with the universal, main r-process pattern that
does not vary from star to star and agrees with the scaled solar
r-process pattern to within measurement uncertainties. If
observational uncertainties are not the reason, this suggests that
the origin of the heavy neutron-capture elements in the r$+$s star
is more complex than previously expected. A possible solution is the
s/r neutron-capture process suggested by \citet{Lugaro2012ApJ},
which is assumed be a single process with features similar to, or a
addition of, the s- and the r-process. If this is true, it is
expected to produce the positive correlations between Ba and Eu
abundances in r$+$s stars, and maybe r-process patterns different
from that of the Sun. However, this hypothesis still needs
theoretical confirmation. In addition, we also cannot exclude the
scenario in which the binary system formed from a gas cloud that was
enriched with r-process material. However, this would imply that the
enrichment event would have resulted in an abundance pattern that at
least in some cases is different from the r-process pattern seen in
the Sun and r-II stars.

From the C/N ratio of $14$ observed in {\SampleStar}, it can be
excluded with high confidence that HBB occurred in its former AGB
companion. According to the evolutionary models of
\citet{Karakas2007PASA}, the former massive companion probably had
an initial mass of less than $3\,\mathrm{M}_\odot$.

The enhanced sodium and magnesium abundances of the star can be
fitted well by the AGB model of \citet{Bisterzo2010MNRAS}, who
highlighted that the primary $^{22}$Ne mainly acted as a neutron
poison in the $^{13}$C-pocket of AGB stars with low mass and
metallicity, which could directly result in a significant production
of Na and Mg. Because $\mathrm{[Mg/Fe]}=0.41$ of {\SampleStar} is
very similar to the value seen in other field stars, a common origin
of Mg, that is pre-enriched by SN~II
\citep{Gehren2006A&A,Andrievsky2010A&A}, cannot be excluded.

The low aluminum abundance also supports the idea of a low-mass AGB
companion of {\SampleStar}, which is consistent with the results
obtained from the model of \citet{Bisterzo2010MNRAS}. The light
elements from calcium to zinc in {\SampleStar} agree well with the
scaled abundance distribution of these elements seen in CS
22892$-$052. This indicates that these elements originate from an
ISM that was already well mixed at the time when these two stars
formed.

\begin{acknowledgements}

  We heartly thank the anonymous referee for positive and constructive comments
  which helped to improve this paper greatly. We thank Dr. Astrid Peter for
  her great help for improving the English expression. W.Y.C. would like to
  thank C. J. Hansen, J. Ren, E. Caffau, L. Sbordone, H.-G. Ludwig,
  K. Andreas, and G. Zhao for their friendly help. This work is supported by Deutsche Forschungsgemeinschaft
  through grant CH 214/5-1 and Sonderforschungsbereich SFB881 ``The
  Milky Way System'' (subproject A5), as well as by the Global Networks
  Program of Universit\"{a}t Heidelberg. W.Y. Cui is also supported by the
  National Natural Science Foundation of China under grant 11003002,
  U1231119, 11273011 and 11021504, the Science Foundation of Hebei Normal University
  under grants L2007B07 and L2009Z04, the Natural Science Foundation of
  Hebei Province under grants A2011205102, and the Program
  for Excellent Innovative Talents in University of Hebei Province under
  grant CPRC034. We made use of model atmosphere from the MARCS library,
  and the NIST and VALD databases.

\end{acknowledgements}

\bibliographystyle{aa}
\bibliography{reference}


\Online

\longtab{5}{
\begin{longtable}[]{lccrrl}
 \caption{\label{Tab:EWlinelist} Line data, equivalent widths, and abundances from the analysis of
   {\SampleStar}. The most
important atomic and molecular data, wavelength $\lambda$,
excitation potential $\chi$, $\log gf$, equivalent width
$W_{\lambda}$, and abundances $\log\epsilon$ are listed.}
\\ \hline
    Species & $\lambda$ & $\chi$ &
   $\mathrm{log}gf$ & $W_{\lambda}$& $\log\epsilon$ \\
   &({\AA})&(eV)&&({m\AA})&\\
   \hline \\

\endfirsthead
\caption{continued.}\\
\hline  \\
    Species & $\lambda$ & $\chi$ &
   $\mathrm{log}gf$ & $W_{\lambda}$& $\log\epsilon$ \\
   &({\AA})&(eV)&&({m\AA})& \\\hline
  \\
\endhead
\hline
\endfoot
\hline 
\endlastfoot
  Mg I   &  5172.684  &  2.710  &  -0.380  &  169.9  &5.368\\
  Mg I   &  5183.604  &  2.720  &  -0.158  &  204.6  &5.540\\
  Si I   &  5948.541  &  5.082  &  -1.230  &   16.3  &5.980\\
  Ca I   &  5265.556  &  2.520  &  -0.260  &   40.5  &4.584\\
  Ca I   &  5349.465  &  2.710  &  -0.310  &   14.5  &4.229\\
  Ca I   &  5581.965  &  2.520  &  -0.710  &   16.6  &4.489\\
  Ca I   &  5588.749  &  2.520  &   0.210  &   53.0  &4.318\\
  Ca I   &  5590.114  &  2.520  &  -0.710  &   12.1  &4.329\\
  Ca I   &  5601.277  &  2.520  &  -0.690  &   26.4  &4.724\\
  Ca I   &  5857.451  &  2.930  &   0.230  &   28.6  &4.277\\
  Ca I   &  6102.723  &  1.880  &  -0.790  &   28.6  &4.163\\
  Ca I   &  6122.217  &  1.890  &  -0.320  &   63.4  &4.309\\
  Ca I   &  6162.173  &  1.900  &  -0.090  &   79.0  &4.353\\
  Ca I   &  6439.075  &  2.520  &   0.470  &   68.6  &4.291\\
  Sc II  &  5031.021  &  1.360  &  -0.400  &   59.5  &1.201\\
  Sc II  &  5526.790  &  1.770  &   0.030  &   51.3  &1.055\\
  Sc II  &  5657.896  &  1.510  &  -0.600  &   35.9  &1.133\\
  Ti I   &  4981.731  &  0.840  &   0.500  &   56.7  &2.925\\
  Ti I   &  4991.065  &  0.840  &   0.380  &   57.2  &3.053\\
  Ti I   &  4999.503  &  0.830  &   0.250  &   64.1  &3.290\\
  Ti I   &  5014.187  &  0.000  &  -1.220  &   40.1  &3.437\\
  Ti I   &  5014.276  &  0.810  &   0.110  &   40.1  &2.998\\
  Ti I   &  5014.187  &  0.000  &  -1.220  &   34.8  &3.341\\
  Ti I   &  5014.276  &  0.810  &   0.110  &   34.8  &2.902\\
  Ti I   &  5192.969  &  0.020  &  -1.010  &   31.4  &3.075\\
  Ti I   &  5210.385  &  0.050  &  -0.880  &   25.1  &2.845\\
  Ti II  &  3500.340  &  0.122  &  -2.020  &   93.7  &3.081\\
  Ti II  &  3504.896  &  1.892  &   0.180  &  116.8  &3.407\\
  Ti II  &  3510.845  &  1.893  &   0.140  &   92.5  &2.843\\
  Ti II  &  4865.612  &  1.116  &  -2.810  &   22.6  &3.079\\
  Ti II  &  5185.913  &  1.893  &  -1.370  &   37.4  &2.772\\
  Ti II  &  5188.680  &  1.582  &  -1.050  &   82.0  &2.860\\
  Ti II  &  5226.543  &  1.566  &  -1.230  &   72.6  &2.850\\
  Ti II  &  5336.771  &  1.582  &  -1.630  &   50.1  &2.890\\
  Ti II  &  5381.015  &  1.566  &  -1.970  &   32.5  &2.908\\
  Cr I   &  5409.772  &  1.030  &  -0.720  &   35.9  &3.145 \\
  Fe I   &  3445.149  &  2.200  &  -0.540  &   96.8  &5.723 \\
  Fe I   &  3490.574  &  0.050  &  -1.110  &  144.7  &5.219 \\
  Fe I   &  3497.841  &  0.110  &  -1.550  &  128.2  &5.370 \\
  Fe I   &  3765.539  &  3.240  &   0.480  &   83.0  &5.049 \\
  Fe I   &  3767.192  &  1.010  &  -0.390  &  140.5  &4.979 \\
  Fe I   &  3787.880  &  1.010  &  -0.860  &  114.5  &4.848 \\
  Fe I   &  4871.318  &  2.870  &  -0.360  &   77.1  &5.133 \\
  Fe I   &  4872.138  &  2.880  &  -0.570  &   71.8  &5.248 \\
  Fe I   &  4891.492  &  2.850  &  -0.110  &   88.3  &5.094 \\
  Fe I   &  4903.310  &  2.880  &  -0.930  &   37.3  &4.982 \\
  Fe I   &  4918.994  &  2.870  &  -0.340  &   72.8  &5.022 \\
  Fe I   &  4939.687  &  0.860  &  -3.340  &   51.2  &5.419 \\
  Fe I   &  4994.130  &  0.920  &  -3.080  &   47.3  &5.155 \\
  Fe I   &  5001.864  &  3.880  &   0.010  &   35.1  &5.071 \\
  Fe I   &  5006.119  &  2.830  &  -0.620  &   60.5  &5.020 \\
  Fe I   &  5041.756  &  1.490  &  -2.200  &   76.0  &5.419 \\
  Fe I   &  5049.820  &  2.280  &  -1.360  &   84.8  &5.627 \\
  Fe I   &  5051.635  &  0.920  &  -2.800  &   67.6  &5.221 \\
  Fe I   &  5068.766  &  2.940  &  -1.040  &   51.5  &5.398 \\
  Fe I   &  5074.748  &  4.220  &  -0.200  &   30.8  &5.553 \\
  Fe I   &  5151.911  &  1.010  &  -3.320  &   66.3  &5.807 \\
  Fe I   &  5166.282  &  0.000  &  -4.200  &   33.6  &4.994 \\
  Fe I   &  5171.596  &  1.490  &  -1.790  &   77.1  &5.017 \\
  Fe I   &  5191.455  &  3.040  &  -0.550  &   71.6  &5.370 \\
  Fe I   &  5192.344  &  3.000  &  -0.420  &   70.5  &5.175 \\
  Fe I   &  5194.942  &  1.560  &  -2.090  &   58.1  &5.041 \\
  Fe I   &  5216.274  &  1.610  &  -2.150  &   46.4  &4.955 \\
  Fe I   &  5225.526  &  0.110  &  -4.790  &   12.8  &5.176 \\
  Fe I   &  5232.940  &  2.940  &  -0.060  &   80.2  &4.935 \\
  Fe I   &  5254.955  &  0.110  &  -4.760  &   19.3  &5.352 \\
  Fe I   &  5266.555  &  3.000  &  -0.390  &   61.4  &4.973 \\
  Fe I   &  5269.537  &  0.860  &  -1.320  &  125.4  &4.931 \\
  Fe I   &  5281.790  &  3.040  &  -0.830  &   33.6  &4.962 \\
  Fe I   &  5283.621  &  3.240  &  -0.520  &   44.8  &5.074 \\
  Fe I   &  5302.302  &  3.280  &  -0.880  &   42.2  &5.430 \\
  Fe I   &  5307.361  &  1.610  &  -2.990  &   15.1  &5.111 \\
  Fe I   &  5324.179  &  3.210  &  -0.240  &   64.9  &5.112   \\
  Fe I   &  5328.039  &  0.920  &  -1.470  &  126.1  &5.152   \\
  Fe I   &  5328.532  &  1.560  &  -1.850  &   81.7  &5.231   \\
  Fe I   &  5339.929  &  3.270  &  -0.720  &   32.7  &5.080   \\
  Fe I   &  5369.962  &  4.370  &   0.540  &   34.9  &5.043   \\
  Fe I   &  5371.490  &  0.960  &  -1.650  &  117.5  &5.163   \\
  Fe I   &  5383.369  &  4.310  &   0.640  &   42.2  &5.016   \\
  Fe I   &  5389.479  &  4.420  &  -0.410  &    6.6  &5.158   \\
  Fe I   &  5393.168  &  3.240  &  -0.910  &   42.6  &5.419   \\
  Fe I   &  5397.128  &  0.920  &  -1.990  &   94.5  &4.918   \\
  Fe I   &  5405.775  &  0.990  &  -1.840  &   96.4  &4.888   \\
  Fe I   &  5424.068  &  4.320  &   0.520  &   43.4  &5.167   \\
  Fe I   &  5429.697  &  0.960  &  -1.880  &  101.9  &5.016   \\
  Fe I   &  5434.524  &  1.010  &  -2.120  &   85.5  &4.951   \\
  Fe I   &  5446.917  &  0.990  &  -1.910  &   94.8  &4.917   \\
  Fe I   &  5455.609  &  1.010  &  -2.090  &  111.2  &5.494   \\
  Fe I   &  5497.516  &  1.010  &  -2.850  &   57.1  &5.148   \\
  Fe I   &  5501.465  &  0.960  &  -3.050  &   55.2  &5.260   \\
  Fe I   &  5506.779  &  0.990  &  -2.800  &   56.2  &5.060   \\
  Fe I   &  5569.618  &  3.420  &  -0.540  &   47.5  &5.321   \\
  Fe I   &  5572.842  &  3.400  &  -0.310  &   54.5  &5.191   \\
  Fe I   &  5576.089  &  3.430  &  -1.000  &   24.0  &5.334   \\
  Fe I   &  5586.756  &  3.370  &  -0.140  &   54.7  &4.991   \\
  Fe I   &  5615.644  &  3.330  &  -0.140  &   61.6  &5.066   \\
  Fe I   &  6136.615  &  2.450  &  -1.400  &   46.8  &5.079   \\
  Fe I   &  6137.692  &  2.590  &  -1.400  &   37.4  &5.068   \\
  Fe I   &  6191.558  &  2.430  &  -1.420  &   32.8  &4.824   \\
  Fe I   &  6213.430  &  2.220  &  -2.480  &   15.5  &5.232   \\
  Fe I   &  6219.281  &  2.200  &  -2.430  &   17.4  &5.219   \\
  Fe I   &  6230.723  &  2.560  &  -1.280  &   44.1  &5.029   \\
  Fe I   &  6252.555  &  2.400  &  -1.690  &   34.3  &5.086   \\
  Fe I   &  6393.601  &  2.430  &  -1.580  &   37.7  &5.065   \\
  Fe I   &  6400.001  &  3.600  &  -0.520  &   41.5  &5.354   \\
  Fe I   &  6421.351  &  2.280  &  -2.030  &   25.6  &5.108   \\
  Fe I   &  6430.846  &  2.180  &  -2.010  &   35.5  &5.177   \\
  Fe I   &  6494.980  &  2.400  &  -1.270  &   51.1  &4.947   \\
  Fe II  &  5197.577  &  3.230  &  -2.230  &   46.6  &5.156   \\
  Fe II  &  5234.625  &  3.220  &  -2.150  &   49.3  &5.109   \\
  Fe II  &  5325.553  &  3.220  &  -3.220  &    7.2  &5.080   \\
  Fe II  &  6247.557  &  3.890  &  -2.330  &   11.2  &5.082   \\
  Fe II  &  6432.680  &  2.890  &  -3.710  &    9.1  &5.274   \\
  Fe II  &  6456.383  &  3.900  &  -2.080  &   25.4  &5.268   \\
  Zn I   &  4722.153  &  4.030  &  -0.338  &   14.0  &2.522  \\
  Zn I   &  4810.528  &  4.078  &  -0.137  &   25.1  &2.478  \\

\hline

\end{longtable}
}

\longtab{6}{
\begin{longtable}[]{lccrrl}
 \caption{\label{Tab:synthlinelist}Line data and abundances
  from the analysis of {\SampleStar}. The most
important atomic data, wavelength $\lambda$, excitation potential
$\chi$, and $\log gf$ are listed.}
\\ \hline
    Species & $\lambda$ & $\chi$ &
   $\mathrm{log}gf$ & $\log\epsilon$ & Ref. \\
   &({\AA})&(eV)&& &\\
   \hline \\

\endfirsthead
\caption{continued.}\\
\hline  \\
    Species & $\lambda$ & $\chi$ &
   $\mathrm{log}gf$ & $\log\epsilon$ & Ref. \\
   &({\AA})&(eV)&& &\\\hline
  \\
\endhead
\hline
\endfoot
\hline 
\endlastfoot
   Li I  &  6707.761  &   0.00  &  -0.009  &   0.90    &  VALD \\
   Li I  &  6707.912  &  0.00  &  -0.309  &  0.90      &  VALD \\
   Be II &  3130.420  &  0.00  &  -0.168  &  $<$-2.86  &  VALD  \\
   Be II &  3131.065  &  0.00  &  -0.468  &  $<$-2.92  &  VALD  \\
   Na I &  5889.951  &  0.00  &   0.117  &   4.56   &  VALD  \\
   Na I &  5895.924  &  0.00  &  -0.184  &   4.68   &  VALD  \\
   Mg I &  3329.919  &  2.71  &  -1.930  &   5.47   &  VALD  \\
   Mg I &  3332.146  &  2.71  &  -1.450  &   5.58   &  VALD  \\
   Mg I &  3336.674  &  2.72  &  -1.230  &   5.53   &  VALD  \\
   Mg I &  3838.290  &  2.72  &  -1.530  &   5.53   &  VALD  \\
   Mg I &  3878.306  &  4.35  &  -0.457  &   5.51   &  VALD  \\
   Mg I &  3903.859  &  4.35  &  -0.511  &   5.57   &  VALD  \\
   Mg I &  4702.991  &  4.35  &  -0.666  &   5.53   &  VALD  \\
   Mg I &  5528.405  &  4.35  &  -0.620  &   5.50   &  VALD  \\
   Al I &  3944.006  &  0.00  &  -0.623  &   3.11   &  VALD  \\
   Al I &  3961.520  &  0.01  &  -0.323  &   3.02   &  VALD  \\
   Si I &  3905.523  &  1.91  &  -0.743  &   4.57   &  VALD  \\
   Si I &  4102.936  &  1.91  &  -2.827  &   4.52   &  VALD  \\
   V  II &  3593.327  &  1.13  &  -0.509  &   1.57   &  VALD  \\
   V  II &  3727.343  &  1.69  &  -0.231  &   1.56   &  VALD  \\
   V  II &  3732.750  &  1.57  &  -0.354  &   1.54   &  VALD  \\
   Cr I &  3578.686  &  0.00  &   0.409  &   3.14   &  VALD  \\
   Cr I &  3593.485  &  0.00  &   0.307  &   3.19   &  VALD  \\
   Cr I &  3605.329  &  0.00  &   0.197  &   3.14   &  VALD  \\
   Cr I &  4254.336  &  0.00  &  -0.114  &   3.13   &  VALD  \\
   Cr I &  4274.797  &  0.00  &  -0.231  &   3.16   &  VALD  \\
   Cr I &  4289.717  &  0.00  &  -0.361  &   3.13   &  VALD  \\
   Cr I &  4344.501  &  1.00  &  -0.550  &   3.12   &  VALD  \\
   Cr I &  4351.811  &  1.03  &  -0.440  &   3.11   &  VALD  \\
   Cr I &  5204.511  &  0.94  &  -0.208  &   3.21   &  VALD  \\
   Cr I &  5206.037  &  0.94  &   0.019  &   3.14   &  VALD  \\
   Cr I &  5208.425  &  0.94  &   0.158  &   3.12   &  VALD  \\
   Cr I &  5264.153  &  0.97  &  -1.290  &   3.19   &  VALD  \\
   Cr I &  5296.691  &  0.98  &  -1.400  &   3.23   &  VALD  \\
   Cr I &  5298.272  &  0.98  &  -1.150  &   3.20   &  VALD  \\
   Cr I &  5345.796  &  1.00  &  -0.980  &   3.14   &  VALD  \\
   Cr I &  5348.315  &  1.00  &  -1.290  &   3.20   &  VALD  \\
   Cr I &  5409.784  &  1.03  &  -0.720  &   3.15   &  VALD  \\
   Mn I &  4030.753  &  0.00  &  -0.470  &   2.59   &  VALD  \\
   Mn I &  4033.062  &  0.00  &  -0.618  &   2.67   &  VALD  \\
   Mn I &  4034.483  &  0.00  &  -0.811  &   2.56   &  VALD  \\
   Mn II &  3438.971  &  1.17  &  -2.100  &   2.64   &  VALD  \\
   Mn II &  3441.985  &  1.78  &  -0.360  &   2.60   &  VALD  \\
   Mn II &  3460.315  &  1.81  &  -0.640  &   2.47   &  VALD  \\
   Mn II &  3482.904  &  1.83  &  -0.840  &   2.62   &  VALD  \\
   Mn II &  3488.675  &  1.85  &  -0.950  &   2.66   &  VALD  \\
   Mn II &  3495.833  &  1.86  &  -1.300  &   2.68   &  VALD  \\
   Mn II &  3496.807  &  1.83  &  -1.790  &   2.63   &  VALD  \\
   Mn II &  3497.525  &  1.85  &  -1.430  &   2.58   &  VALD  \\
   Co I &  3518.346  &  1.05  &   0.070  &   2.34   &  VALD  \\
   Co I &  3995.302  &  0.92  &  -0.220  &   2.37   &  VALD  \\
   Ni I &  4980.166  &  3.61  &   0.070  &   3.98   &  VALD  \\
   Ni I &  5035.357  &  3.64  &   0.290  &   4.02   &  VALD  \\
   Ni I &  5137.070  &  1.68  &  -1.990  &   3.97   &  VALD  \\
   Ni I &  5476.900  &  1.83  &  -0.890  &   4.08   &  VALD  \\
   Ni I &  5709.539  &  1.68  &  -2.170  &   3.99   &  VALD  \\
   Cu I &  3247.537  &  0.00  &  -0.062  &   0.66   &  VALD  \\
   Cu I &  3273.954  &  0.00  &  -0.359  &   0.73   &  VALD  \\
   Sr II &  4077.709  &  0.00  &   0.167  &   0.37   &  VALD  \\
   Sr II &  4215.519  &  0.00  &  -0.145  &   0.36   &  VALD  \\
   Y  II &  3601.919  &  0.10  &  -0.180  &   0.00   &  VALD  \\
   Y  II &  3774.331  &  0.13  &   0.210  &   0.09   &  VALD  \\
   Y  II &  3950.352  &  0.10  &  -0.490  &   0.16   &  VALD  \\
   Y  II &  3982.594  &  0.13  &  -0.490  &   0.06   &  VALD  \\
   Y  II &  4374.935  &  0.41  &   0.160  &   0.11   &  VALD  \\
   Y  II &  4422.591  &  0.10  &  -1.270  &   0.12   &  VALD  \\
   Y  II &  4854.863  &  0.99  &  -0.380  &   0.10   &  VALD  \\
   Y  II &  4883.684  &  1.08  &   0.070  &   0.04   &  VALD  \\
   Y  II &  4900.120  &  1.03  &  -0.090  &   0.13   &  VALD  \\
   Y  II &  5087.416  &  1.08  &  -0.170  &   0.10   &  VALD  \\
   Y  II &  5200.406  &  0.99  &  -0.570  &   0.15   &  VALD  \\
   Y  II &  5205.724  &  1.03  &  -0.340  &   0.10   &  VALD  \\
   Y  II &  5662.925  &  1.94  &   0.160  &   0.10   &  VALD  \\
   Zr II &  3457.548  &  0.56  &  -0.530  &   0.95   &  VALD  \\
   Zr II &  3481.137  &  0.80  &   0.165  &   0.90   &  VALD  \\
   Zr II &  3499.560  &  0.41  &  -0.810  &   0.90   &  VALD  \\
   Zr II &  3551.939  &  0.09  &  -0.310  &   1.08   &  VALD  \\
   Zr II &  3611.889  &  1.74  &   0.450  &   1.07   &  VALD  \\
   Zr II &  3614.765  &  0.36  &  -0.252  &   0.90   &  VALD  \\
   Zr II &  3630.004  &  0.36  &  -1.110  &   1.09   &  VALD  \\
   Zr II &  3668.432  &  0.41  &  -1.138  &   0.94   &  VALD  \\
   Zr II &  3714.794  &  0.53  &  -0.930  &   0.90   &  VALD  \\
   Zr II &  3751.606  &  0.97  &   0.012  &   1.02   &  VALD  \\
   Zr II &  3766.795  &  0.41  &  -0.812  &   1.09   &  VALD  \\
   Zr II &  3991.152  &  0.76  &  -0.252  &   0.91   &  VALD  \\
   Zr II &  3998.954  &  0.56  &  -0.387  &   1.00   &  VALD  \\
   Zr II &  4149.217  &  0.80  &  -0.030  &   0.94   &  VALD  \\
   Zr II &  4208.977  &  0.71  &  -0.460  &   0.92   &  VALD  \\
   Zr II &  4496.962  &  0.71  &  -0.890  &   0.94   &  VALD  \\
   Nb II &  3130.780  &  0.44  &   0.410  &  -0.00   &  VALD  \\
   Nb II &  3163.398  &  0.38  &   0.260  &   0.40   &  VALD  \\
   Ba II &  3891.776  &  2.51  &   0.280  &   1.63   &  VALD  \\
   Ba II* &  4130.700  &  2.72  &   0.560  &   1.71   & McW98  \\
   Ba II &  4524.925  &  2.51  &  -0.360  &   1.69   &  VALD  \\
   Ba II* &  4554.000  &  0.00  &   0.170  &   1.79   & McW98  \\
   Ba II &  4899.929  &  2.72  &  -0.080  &   1.70   &  VALD  \\
   Ba II* &  4934.100  &  0.00  &  -0.150  &   1.66   & McW98  \\
   Ba II* &  5853.700  &  0.60  &  -1.010  &   1.67   & McW98  \\
   Ba II* &  6141.695  &  0.70  &  -0.070  &   1.68   & McW98  \\
   Ba II &  6496.897  &  0.60  &  -0.377  &   1.68   &  VALD  \\
   La II* &  4808.996  &  0.23  &  -1.400  &   0.26   & Law01a,Ivan06 \\
   La II &  4899.915  &  0.00  &  -0.921  &   0.06   & Law01a \\
   La II &  4920.976  &  0.13  &  -0.730  &   0.26   & Law01a \\
   La II &  4921.776  &  0.24  &  -0.450  &   0.26   & Law01a \\
   La II* &  4970.386  &  0.32  &  -1.160  &   0.16   & Law01a,Ivan06 \\
   La II* &  4986.819  &  0.17  &  -1.300  &   0.06   & Law01a,Ivan06 \\
   La II* &  4999.461  &  0.40  &  -0.770  &   0.32   & Law01a,Ivan06 \\
   La II* &  5114.559  &  0.23  &  -1.032  &   0.30   & Law01a,Ivan06 \\
   La II* &  5122.988  &  0.32  &  -0.850  &   0.31   & Law01a,Ivan06 \\
   La II &  5259.379  &  0.17  &  -1.950  &   0.30   & Law01a \\
   La II &  5290.818  &  0.00  &  -1.650  &   0.35   & Law01a \\
   La II* &  5303.528  &  0.32  &  -1.350  &   0.30   & Law01a,Ivan06 \\
   La II* &  5482.268  &  0.00  &  -2.230  &   0.33   & Law01a,Ivan06 \\
   Ce II &  3655.844  &  0.32  &   0.233  &   0.15   &  VALD  \\
   Ce II &  4042.581  &  0.50  &   0.070  &   0.10   &  VALD  \\
   Ce II &  4053.503  &  0.00  &  -0.460  &   0.16   &  VALD  \\
   Ce II &  4120.827  &  0.32  &  -0.130  &   0.10   &  VALD  \\
   Ce II &  4137.645  &  0.52  &   0.246  &   0.10   &  VALD  \\
   Ce II &  4186.594  &  0.86  &   0.813  &   0.15   &  VALD  \\
   Ce II &  4222.597  &  0.12  &  -0.301  &   0.18   &  VALD  \\
   Ce II &  4364.653  &  0.50  &   0.070  &   0.15   &  VALD  \\
   Ce II &  4479.361  &  0.56  &  -0.480  &   0.20   &  VALD  \\
   Ce II &  4486.909  &  0.30  &  -0.260  &   0.20   &  VALD  \\
   Ce II &  4523.075  &  0.52  &  -0.030  &   0.15   &  VALD  \\
   Ce II &  4562.359  &  0.48  &   0.230  &   0.13   &  VALD  \\
   Ce II &  4572.278  &  0.68  &   0.290  &   0.22   &  VALD  \\
   Ce II &  5187.458  &  1.21  &   0.300  &   0.19   &  VALD  \\
   Ce II &  5274.229  &  1.04  &   0.300  &   0.13   &  VALD  \\
   Pr II &  3908.428  &  0.00  &   0.019  &  -0.13   &  VALD  \\
   Pr II &  4100.717  &  0.55  &   0.572  &  -0.26   &  VALD  \\
   Pr II* &  4143.120  &  0.37  &   0.609  &  -0.13   &  Ivar01, Sned09  \\
   Pr II* &  4222.950  &  0.05  &   0.271  &  -0.18   &  Ivar01, Sned09  \\
   Pr II* &  5173.910  &  0.97  &   0.384  &  -0.19   &  Ivar01, Sned09  \\
   Pr II* &  5220.108  &  0.80  &   0.298  &  -0.18   &  Ivar01, Sned09  \\
   Pr II* &  5259.728  &  0.63  &   0.114  &  -0.14   &  Ivar01, Sned09  \\
   Pr II* &  5322.772  &  0.48  &  -0.319  &  -0.23   &  Ivar01, Sned09  \\
   Nd II &  5249.576  &  0.98  &   0.094  &   0.63   &  VALD  \\
   Nd II &  5250.812  &  0.75  &  -0.618  &   0.72   &  VALD  \\
   Nd II &  5255.506  &  0.20  &  -0.697  &   0.72   &  VALD  \\
   Nd II &  5273.427  &  0.68  &  -0.185  &   0.71   &  VALD  \\
   Nd II &  5293.163  &  0.82  &  -0.144  &   0.70   &  VALD  \\
   Nd II &  5303.200  &  0.38  &  -1.324  &   0.71   &  VALD  \\
   Nd II &  5311.453  &  0.99  &  -0.437  &   0.72   &  VALD  \\
   Nd II &  5319.815  &  0.55  &  -0.152  &   0.67   &  VALD  \\
   Nd II &  5336.532  &  0.55  &  -1.074  &   0.72   &  VALD  \\
   Nd II &  5356.967  &  1.26  &  -0.248  &   0.71   &  VALD  \\
   Nd II &  5361.467  &  0.68  &  -0.482  &   0.71   &  VALD  \\
   Nd II &  5385.888  &  0.74  &  -0.860  &   0.72   &  VALD  \\
   Nd II &  5485.696  &  1.26  &  -0.284  &   0.65   &  VALD  \\
   Nd II &  5594.416  &  1.12  &  -0.279  &   0.73   &  VALD  \\
   Nd II &  5688.518  &  0.99  &  -0.404  &   0.72   &  VALD  \\
   Nd II &  5702.238  &  0.75  &  -0.744  &   0.72   &  VALD  \\
   Nd II &  5708.271  &  0.86  &  -0.581  &   0.70   &  VALD  \\
   Sm II &  4318.927  &  0.28  &  -0.268  &  -0.34   &  VALD  \\
   Sm II &  4420.524  &  0.33  &  -0.383  &  -0.32   &  VALD  \\
   Sm II &  4537.941  &  0.49  &  -0.230  &  -0.28   &  VALD  \\
   Sm II &  4543.943  &  0.33  &  -0.577  &  -0.29   &  VALD  \\
   Sm II &  4815.805  &  0.19  &  -0.775  &  -0.35   &  VALD  \\
   Eu II* &  4129.628  &  0.00  &  0.210  &  -0.29   & Law01c \\
   Eu II* &  4204.908  &  0.00  &  0.220  &  -0.33   & Law01c \\
   Gd II &  3646.196  &  0.24  &   0.328  &  -0.10   &  VALD  \\
   Gd II &  4130.366  &  0.73  &  -0.090  &  -0.20   &  VALD  \\
   Gd II &  4251.731  &  0.38  &  -0.365  &  -0.20   &  VALD  \\
   Gd II &  4325.557  &  1.37  &   0.772  &  -0.23   &  VALD  \\
   Gd II &  4327.151  &  0.35  &  -0.37  &  -0.22   &  VALD  \\
   Tb II &  3509.144  &  0.00  &   0.700  &  -0.98   &  law01b,law01d  \\
   Dy II &  3694.810  &  0.10  &   0.000  &  -0.21   &  VALD  \\
   Dy II &  3836.505  &  0.54  &  -0.025  &  -0.21   &  VALD  \\
   Dy II &  3898.528  &  0.59  &   0.397  &  -0.19   &  VALD  \\
   Dy II &  3944.681  &  0.00  &   0.100  &  -0.11   &  VALD  \\
   Dy II &  3978.561  &  0.93  &   0.360  &  -0.13   &  VALD  \\
   Dy II &  4077.966  &  0.10  &   0.010  &  -0.21   &  VALD  \\
   Dy II &  4957.348  &  0.00  &  -1.010  &  -0.16   &  VALD  \\
   Er II &  3692.649  &  0.05  &   0.138  &  -0.23   &  VALD  \\
   Yb* II &  3289.367  &  0.00  &  -0.052  &   0.39   &  BDM98,Sned09  \\
   Yb* II &  3694.192  &  0.00  &  -0.320  &   0.97   &  BDM98,Sned09  \\
   Lu II &  3507.380  &  0.00  &  -1.160  &  -0.91   &  VALD  \\
   Lu II &  4994.126  &  1.54  &  -1.320  &  -1.05   &  VALD  \\
   Lu II &  5476.676  &  1.76  &  -0.276  &  -1.05   &  VALD  \\
   Hf II &  3478.980  &  2.16  &   0.280  &   0.18   &  VALD  \\
   Hf II &  3479.289  &  0.38  &  -1.040  &   0.07   &  VALD  \\
   Hf II &  3505.219  &  1.04  &  -0.080  &   0.06   &  VALD  \\
   Hf II &  3535.549  &  0.61  &  -0.540  &   0.03   &  VALD  \\
   Hf II &  3569.034  &  0.79  &  -0.400  &   0.08   &  VALD  \\
   Hf II &  3597.394  &  1.89  &  -0.030  &   0.00   &  VALD  \\
   Hf II &  3644.352  &  0.79  &  -0.480  &   0.13   &  VALD  \\
   Hf II &  3661.045  &  1.87  &  -0.320  &   0.10   &  VALD  \\
   Hf II &  3699.731  &  1.67  &  -0.300  &   0.08   &  VALD  \\
   Hf II &  3701.156  &  2.16  &   0.330  &   0.18   &  VALD  \\
   Hf II &  3719.276  &  0.61  &  -0.870  &   0.08   &  VALD  \\
   Hf II &  3737.869  &  2.34  &   0.160  &   0.07   &  VALD  \\
   Pb I &  3683.462  &  0.97  &  -0.460  &   1.93   &  VALD  \\
   Pb I* &  4057.807  &  1.32  &  -0.220  &   1.96   & Van03 \\

\hline

\end{longtable}
\begin{center}
* HFS is included only for one of transitions listed here.
\end{center}
\begin{table}[hb]
\centering
\begin{tabular}{ll}
\multicolumn{1}{l}{BDM98} & \multicolumn{1}{l}{\citet{Biemont1998JPhB};} \\
\multicolumn{1}{l}{McW95} & \multicolumn{1}{l}{\citet{McWilliam1995AJ};} \\
\multicolumn{1}{l}{McW98} & \multicolumn{1}{l}{\citet{McWilliam1998AJ};} \\
\multicolumn{1}{l}{Ivar01} & \multicolumn{1}{l}{\citet{Ivarsson2001PhyS};} \\
\multicolumn{1}{l}{Ivan06} & \multicolumn{1}{l}{\citet{Ivans2006ApJ};} \\
\multicolumn{1}{l}{Law01a} & \multicolumn{1}{l}{\citet{Lawler2001ApJL};} \\
\multicolumn{1}{l}{Law01b} & \multicolumn{1}{l}{ \citet{Lawler2001ApJS};}\\
\multicolumn{1}{l}{Law01c} & \multicolumn{1}{l}{ \citet{Lawler2001ApJ};}\\
\multicolumn{1}{l}{Law01d} & \multicolumn{1}{l}{ \citet{Lawler2001ApJSa};}\\
\multicolumn{1}{l}{Sned09} & \multicolumn{1}{l}{ \citet{Sneden2009ApJS};}\\
\multicolumn{1}{l}{VALD} & \multicolumn{1}{l}{\citet{Kupka1999A&AS};} \\
\multicolumn{1}{l}{Van03}&\multicolumn{1}{l}{\citet{VanEck2003A&A}.}

\end{tabular}
\end{table}
}

\end{document}